\documentclass{ws-procs9x6}

\begin{document}

\title{Dark Matter on Galactic Scales \\(or the Lack Thereof)}

\author{Michael R. Merrifield}

\address{School of Physics \& Astronomy\\
University of Nottingham\\
University Park, Nottingham\\
NG7 2RD, United Kingdom\\
E-mail: michael.merrifield@nottingham.ac.uk}

\maketitle

\abstracts{ This paper presents a brief review of the evidence for
dark matter in the Universe on the scales of galaxies.  In the
interests of critically and objectively testing the dark matter
paradigm on these scales, this evidence is weighed against that from
the only other game in town, modified Newtonian dynamics.  The verdict
is not as clear cut as one might have hoped.}

\section{Introduction}\label{sec:intro}
In the study of dark matter, its properties on the galactic scale are
of abiding interest.  It is, after all, the signature of a galaxy's
dark halo that almost all of the current dark matter experiments are
seeking to detect --- they are trying to interact with the rather
small part of the Milky Way's massive halo that happens to be passing
through the experimenters' laboratories at the moment.  Any
predictions as to detection rates, etc, are therefore dependent on our
understanding the properties of galactic-scale dark matter halos.

Galaxies have also played a pivotal role in the entire development of
the dark matter paradigm.  In one of the founding papers on the
subject,\cite{z37} Zwicky used the galaxies in the Coma Cluster as
test particles to which he applied the virial theorem.  This
calculation led to an unexpectedly high mass for the cluster, and
Zwicky pointed out that if this total were divided amongst the
constituent galaxies, then they would each have an average mass of $5
\times 10^{10} M_\odot$ (where $M_\odot$ is the mass of the Sun),
whereas their average luminosity was estimated to be only $9 \times
10^7 L_\odot$ (where $L_\odot$ is the Sun's luminosity).  Clearly,
such galaxies could not be made up of stars like the Sun, and some
extra contributor to the mass is required.  In this prescient paper,
Zwicky also pointed out that gravitational lensing should provide
another approach to determining the masses of clusters, and that the
rotation curves of individual galaxies --- their speed of rotation as
a function of radius, $v_c(r)$ --- should allow one to probe the
distribution of mass within these systems by equating the centripetal
acceleration to the gravitational acceleration due to the mass.

Early studies of rotation curves were hampered by the fact that the
optical emission lines used to measure Doppler shifts, and hence
rotation speeds, could only be detected in the inner parts of
galaxies.  In general, at these small radii the inferred mass is quite
consistent with the rather uncertain census of luminous
components.\cite{k86} However, the discovery that spiral galaxies are
surrounded by huge disks of atomic hydrogen completely changed the
picture.\cite{k87} Using the Doppler shifts in this gas' 21cm radio
emission, it became clear that the rotation speeds of such galaxies
remain roughly constant out to the largest radii measured, $v_c \sim
{\rm constant}$.  This is certainly not what would have been predicted
on the basis of the known luminous mass, as the measurements were
made at radii well beyond most of these components, so one would have
expected the velocities to have entered a Keplerian decline, with $v_c
\propto r^{-1/2}$.  These incontrovertible observations were
fundamental ingredients in a huge shift of paradigm, taking
astronomers from a position in which evidence for dark matter could be
quietly swept under the cosmic rug to one in which it was recognised
that dark matter seems to dominate the Universe on almost all scales.

Over the last twenty-five years, this recognition has spawned an
entire industry of simulators trying to model the formation of
structure in the Universe through the gravitational collapse of
dark-matter-dominated material, concentrating on the idea that this
mass comprises gravitationally-interacting non-relativistic ``cold
dark matter'' (CDM).  In the early days, these simulations lacked the
resolution to study the properties of individual galaxies, but more
recently very large computer simulations have achieved the dynamic
range necessary to study the formation and properties of individual
galaxies in CDM cosmology.\cite{n04} These simulations now make some
fairly definite and robust predictions as to the distribution of dark
matter on galactic scales.  In particular, they predict that the dark
matter should be distributed such that its density distribution has a
central power-law cusp, $\rho \propto r^{-\gamma}$.  Authors differ
somewhat as to the steepness of this cusp,\cite{n04} but it seems to
have a power-law index of between $\gamma = 1$ and $\gamma = 1.5$.
However, all seem to agree that at large radii the density profile
steepens to a power-law index of $\gamma = 3$.  This all fits
reasonably well with the observed flat rotation curves, which would be
produced by a density profile with $\gamma = 2$, as the measured
rotation curves typically probe intermediate radii between these two
extreme regimes.

The simulations also predict the over-all shapes of the dark halos,
which are typically oblate with a range of shortest-to-longest axis
ratios of $\sim 0.5 \pm 0.2$.\cite{d94} This quantity is hard to
determine observationally --- the rotation curve of a spiral galaxy,
for example, provides a useful measure of the radial distribution of
mass in a galaxy, but not its distribution perpendicular to the plane
of the galaxy's disk.  However, probes of this third dimension do
exist in the form of the orbits of material in merging satellites, the
thickness of the gas layer, and the shapes of hot X-ray emitting gas
halos around elliptical galaxies.  As far as can be ascertained from
these limited data, the observed shapes of halos are consistent with
the CDM predictions.\cite{m04}

Not everything in the CDM simulations fits so well with the
observations, though.  In particular, the repeated merging of
sub-halos that produces galaxies in these simulations does not produce
a simple smooth density distribution.  Instead, the simulations
predict that one should see large amounts of substructure in the form
of small satellite systems around every large galaxy.  Although big
objects like the Milky Way are accompanied by a retinue of smaller
satellites, there seems to be a problem in that CDM simulations
predict far more companions than are observed.\cite{m99} This galaxy
formation picture is also beginning to run into trouble with
observations at high redshift: in this hierarchical scenario, the
largest most massive galaxies should form last, yet observations
indicate that some fraction of very luminous galaxies are already in
place quite early in the Universe's history.\cite{s04}

\section{The Alternative to Dark Matter}\label{sec:MOND}
Although the dark matter paradigm is almost universally accepted in
the astronomical community, it does seem to have its limitations, so
it still pays to step back from it occasionally and try to weigh it up
against competing theories.  In this regard, the only real alternative
is the Modified Newtonian Dynamics or MOND hypothesis.\cite{m83,s02}
In this theory, the Newtonian acceleration due to gravity, $a_n$, is
replaced by an acceleration $a$ that obeys the equation $a_n = a
\times \mu(|a|/a_0)$.  The function $\mu$ is chosen so that $a$ varies
smoothly from $a_n$ when $a \gg a_0$ to $\sqrt{a_n \times a_0}$ when
$a \ll a_0$.  The only free parameter in the theory is the
characteristic acceleration constant $a_0$ at which the transition
occurs.

This modification to Newton's laws neatly explains the flatness of the
outer parts of galaxy rotation curves. At these large radii, Newtonian
acceleration due to gravity is simply $a_n \approx {G M \over r^2}$, where
$M$ is the galaxy's total mass.  Centripetal acceleration here is
sufficiently low that we are in the MOND regime, so we can write
\begin{equation}
{v_c^2 \over r} = a \approx \sqrt{{G M \over r^2} \times a_0}. 
\end{equation}
The $r$ dependence cancels from the equation, so we find that 
\begin{equation}
v_c \approx (G M a_0)^{1/4}, 
\label{eq:vcmond}
\end{equation}
independent of radius.

This amendment to Newton's laws is somewhat {\it ad hoc}, and was
clearly reverse-engineered to deal with the issue of flat rotation
curves.  However, it is quite legitimate to ask whether this solution
is any more arbitrary than postulating the existence of invisible mass
of an unspecified nature to explain away the perceived problem of
galaxies' rotation curves.  It is also worth noting that MOND provides
explanations for various other astronomical phenomena at no extra
charge.  For example, there is observed to be a tight correlation
between asymptotic rotation speeds and luminosities in galaxies, such
that $L \propto v_c^4$.  This proportionality, known as the
Tully--Fisher relation,\cite{tf77} has no particularly fundamental
explanation in standard Newtonian gravity.  However, if all that a
galaxy contains is its luminous stars, so that $L \propto M$, then the
Tully--Fisher relation would follow trivially from
Eq.~(\ref{eq:vcmond}) if MOND were correct.

A further criticism arising from the {\it ad hoc} nature of MOND is
that it has no satisfactory relativistic generalisation.  However,
this situation changed dramatically earlier this year when a fully
relativistic field theory was developed that reduces to Newtonian
dynamics in the low-velocity regime and to MOND at low
accelerations.\cite{b04} This theory for the first time makes a robust
prediction as to what gravitational lensing signature one might expect
from MOND --- another shortcoming often pointed out in the theory ---
and shows that it will be of the same amplitude as that predicted on
the basis of dark matter models.

With this new relativistic formulation, it is also now possible to
carry out fully self-consistent tests of MOND such as performing
simulations of structure formation in a cosmological context, to see
if the theory fits with other astronomical observations as closely as
the CDM simulations.  Although no-one has yet carried out such
simulations, there are already some interesting insights that one can
obtain into the subject.  The longer range nature of gravity in MOND
means that the two-body relaxation time and dynamical friction
timescale are dramatically shorter than in Newtonian
gravity.\cite{c04} One might therefore expect the over-abundance of
substructure found in the CDM simulations (see Sec.~\ref{sec:intro})
to be effectively wiped out in a MOND universe, as such structure
should relax into a smooth distribution on a relatively short
timescale.  However, whether this simple heuristic argument works in
detail will only become clear once some full cosmological MOND
simulations are performed.

\section{Case Studies}\label{sec:cases}
To further test the dark matter paradigm on galactic scales, let us
now turn to a few specific types of galaxy to see how well the data
fit the theory.  There is not much point in using normal spiral
galaxies for such tests, as these systems played a key role in the
development of the dark matter paradigm (and, indeed, of MOND as
well); it would be very surprising if the theory failed to reproduce
the properties that so strongly motivated it in the first place.
Instead, we look to other types of galaxy for some independent
confirmation or refutation of the theory.

\subsection{Low Surface Brightness Galaxies}
Actually, spiral galaxies are rather poor places to test the dark
matter paradigm for other reasons, too.  One of the clear predictions
in CDM is that the dark matter should have a central cusp.  However,
the large amount of luminous matter at small radii in these systems
means that such cusps will have very little impact on the rotation
curves of spiral galaxies.  Further, the gravitational interaction
between luminous and dark matter could well have redistributed the
dark matter, so any primordial central cusp may have been completely
wiped out.

Fortunately, a class of galaxies exists in which these issues should
not arise.  These ``low surface brightness galaxies'' contain a very low
density of luminous material even in the inner parts, so that the
observed dynamics should be dominated by the gravitational forces of
the dark halo at small radii as well as large radii.  Further, the
small amount of luminous matter should not have had much ability to
redistribute the dark matter, so the central cusp should still be
there. 

\begin{figure}[th]
\centerline{\epsfxsize=4.1in\epsfbox{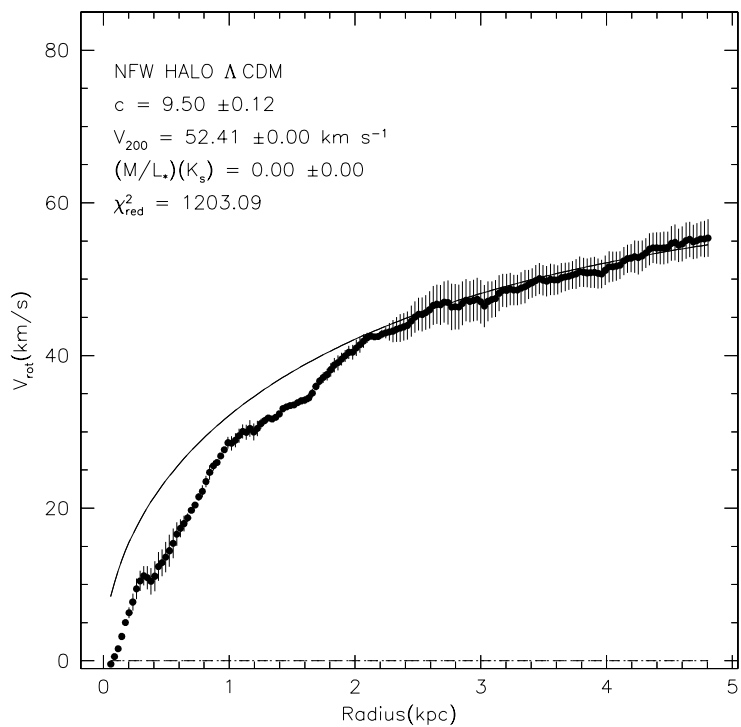}}   
\caption{Rotation curve (circular rotation velocity as a function of
radius) for the low surface brightness galaxy NGC~6822.  The points
show data obtained from the 21cm emission of atomic hydrogen, while
the line gives the best-fit model assuming that the galaxy's mass is
dominated by a centrally-cusped dark matter halo.  (Figure kindly
provided by W.J.G. de Blok.)
\label{fig:LSB}}
\end{figure}

Rather disturbingly, as illustrated in Fig.~\ref{fig:LSB}, the
rotation curves of these systems do not seem to match up to the
predictions of the CDM models.  Although they flatten off to the
constant rotation velocity characteristic of a dark matter halo, they
rise significantly more slowly in their central parts than one would
expect for a system with a centrally-concentrated cusped mass
distribution.\cite{db02}

Once again, MOND seems to do rather better.\cite{db98}  Not only does it
reproduce the observed slowly-rising rotation curves, but in many
cases it can also explain the match up between localized features in
the rotation curves and small-scale structure in the photometry --- if
these low surface brightness galaxies were dominated by dark matter
halos, then the faint features in the luminous light distribution
should have almost no direct impact on the rotation curve.  However,
one has to keep a critical eye on how illustrative examples are
selected: in the case of Ref.~[\refcite{s02}], the highlighted low
surface brightness galaxy in their Fig.~3 shows a beautiful match in
the small-scale structure of the rotation curve and the MOND
predictions, whereas the broader range of cases in their Figs.~4 and 5 show
a number of examples where there is no such correspondence.

\subsection{Elliptical Galaxies}
The other main class of luminous galaxy are the rather dull-looking
elliptical systems.  Studies of the dark matter in these objects have
been hampered for several reasons.  First, they tend to be found in
clusters, so it becomes difficult (or even an issue of semantics) to
separate any dark matter associated with galaxies from that which should
be attributed to the cluster over all.  Second, they lack the
extensive gas disks that surround spiral galaxies.  They therefore do
not have a simple kinematic tracer that allows their masses to be
measured out to large radii.  

For the largest elliptical galaxies, alternative probes such as
gravitational lensing\cite{k01} and X-ray emitting gas\cite{l99} have
been used to infer that these systems have extensive massive halos.
However, such bright galaxies tend to lie at the centres of clusters,
raising the issue of whether the halo belongs to the galaxy or the
cluster.

\begin{figure}[th]
\centerline{\epsfxsize=4.1in\epsfbox{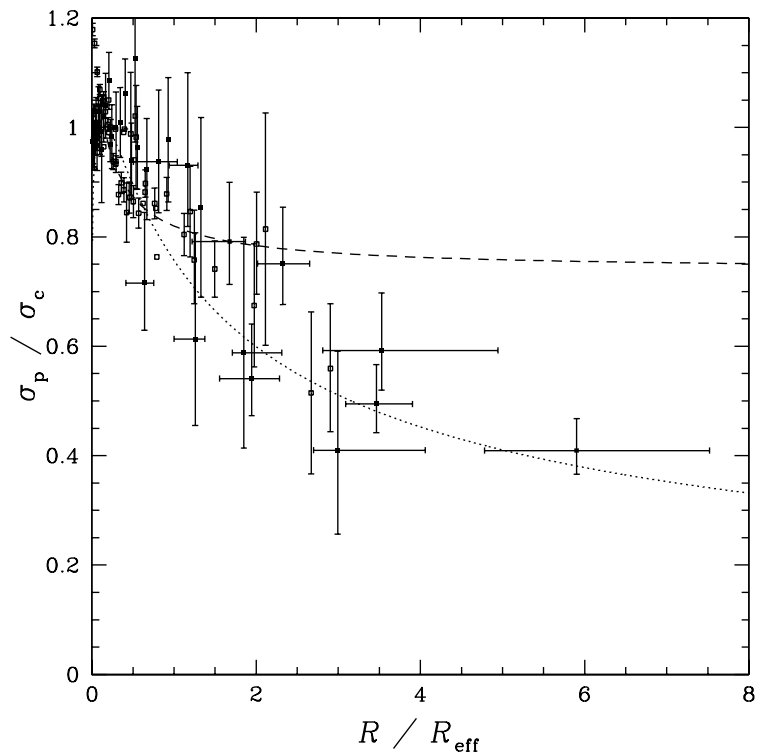}}   
\caption{A compilation of the stellar line-of-sight velocity
dispersion as a function of projected radius for four
intermediate-luminosity elliptical galaxies.  The data for different
galaxies have been normalized by the central velocity dispersion and
the effective radius (the radius within which half of the galaxy's
light appears projected) to make them directly comparable.  The dashed
line shows the predicted trend in this relation if these galaxies were
embedded in massive halos comparable to those found around similar
luminosity spirals.  The dotted line shows the Keplerian decline that
would be found in the absence of a dark halo.
\label{fig:PNS}}
\end{figure}

For more ``normal'' ellipticals, comparable in luminosity to the Milky
Way, not a lot of data have been available.  However, we have recently
developed a new instrument, the Planetary Nebula Spectrograph,
specifically to study such objects.\cite{d02} This instrument detects
and measures the kinematics of planetary nebulae (PNe) in the outer
parts of elliptical galaxies.  Since PNe are simply stars that have
reached the ends of their lives, their kinematics will be
representative of the over-all stellar population (but much easier to
measure due to the bright emission lines in PNe spectra).  The initial
results that we have obtained using this instrument were not at all
what we expected.\cite{r03} As Fig.~\ref{fig:PNS} shows, the random
velocities of the PNe do not stay constant out to large radii, as one
might expect by analogy with the rotation curve of a comparable spiral
galaxy; instead they seem to go into Keplerian decline, as would be
expected in the complete absence of dark matter.

There are a number of possible explanations for this result that would
not involve making as radical a change as to suggest that these
systems are devoid of dark matter.  First, there is an extra degree of
freedom that we did not have to worry about for spiral galaxies: while
the gas in disks around spirals all follows orbits that are close to
circular, we have no {\it a priori} way of knowing whether the stars
in ellipticals (and hence their PNe) follow circular orbits, radial
orbits, or something in between.  Since objects on radial orbits will
move mostly transverse to the line of sight at large radii, one would
expect the measured line-of-sight velocity dispersion to drop for a
system consisting of such objects.  However, it would be a surprising
coincidence if such an extreme collection of orbits managed to mimic
the behaviour of a Keplerian decline; indeed, detailed orbit modelling
indicates that such a solution is not consistent with the
data.\cite{r03} Alternatively, these galaxies could be surrounded by
halos that are so extended that, even at the unprecedentedly large
radii to which we are now looking, the dark matter still does not
dominate the mass.  However, this scenario also seems to conflict with
CDM models, which predict that these moderate-luminosity galaxies
should have dark halos that are more centrally concentrated than their
brighter kin, not less so.\cite{n04b}

Although this result was not what we were expecting, it did not come
as a surprise to the advocates of MOND.  Indeed, Ref.~[\refcite{s02}]
had already made the definite prediction that lower-luminosity
elliptical galaxies should have velocity dispersion profiles that fall
with radius, and subsequent fitting to our data confirmed that the
observed approximately Keplerian decline is what would be expected in
MOND.\cite{m03} Essentially, the absence of any dark matter signature
in these systems arises because elliptical galaxies are more centrally
concentrated than their spiral cousins, which means that the
characteristic accelerations within them stay safely above the MOND
$a_0$ threshold.  Thus, their observed kinematics should be consistent
with ordinary Newtonian physics, with no non-standard dynamics to be
mis-interpreted as dark matter.

\section{Conclusions}\label{sec:conc}
The idea that we live in a Universe dominated by dark matter is so
deeply embedded in most astronomers' World views that it is not
something that many of us ever question.  Nonetheless, all of the
evidence that we currently have for dark matter is highly
circumstantial, so we should at least compare the hypothesis
critically with any alternatives.

In this regard, a comparison between the dark matter hypothesis and
modified Newtonian dynamics, at least on the galactic scale, produces
a rather equivocal answer.  In a number of cases, MOND seems to fit
the observations rather better than dark matter.  There are also
examples where MOND has passed the ultimate test of a scientific
theory by making predictions that differ from the dark matter theory,
which subsequently turn out to be true.  On scales other than the
galactic, perhaps MOND does a little less well: even its proponents
recognise that there seems to be a problem at the size of clusters,
with MOND predicting more mass than can be found within the known
luminous components.\cite{s99} However, until some serious effort is
invested in proper cosmological MOND simulations, we will not be able
to establish the context that will determine quite how serious any
problems might be with this theory.

As things stand, it is principally a matter of aesthetics as to which
idea Occam's Razor favours.  Personally, I would still go for the
dark matter theory: it is astoundingly arrogant to assume that everything
in the Universe should glow in the dark for the benefit of
astronomers, so invoking dark matter appeals to me as a way to remind
us of our own insignificance.  Nonetheless, I recognise the appeal of
a simple modification to the law of gravity over the invocation of an
entirely unknown form of matter, so am unable to draw any definite
conclusions even on this aesthetic issue.

In the context of this particular volume, however, the message is much
more clear cut: the direct laboratory detection of massive particles
from the Milky Way's halo would provide by far the most convincing
confirmation of the whole dark matter paradigm, and would lay this
issue to rest once and for all.

\end{document}